\documentclass[12pt,a4paper]{article}

\usepackage{ulem}
\usepackage{color,amsmath}
\usepackage{amsfonts}
\usepackage{amssymb}
\usepackage{graphicx}
\usepackage{epsfig}
\usepackage{enumerate}
\usepackage{stmaryrd}
\usepackage{varioref,exscale,relsize}
\usepackage{amsthm}
\usepackage{mathrsfs}
\usepackage{amsfonts}
\usepackage{enumerate}
\usepackage{subfig}

\IfFileExists{url.sty}{\usepackage{url}}
                     {\newcommand{\url}{\text}}

\DeclareRobustCommand{\FIN}{%
    \ifmmode 
    \else \leavevmode\unskip\penalty9999 \hbox{}\nobreak\hfill
    \fi
    $\bullet$ \vspace{5mm}}

\begin{document}

\

\begin{center}
{\Large \textbf{Sequential Clustering for Functional Data}}

{ Ana Justel$^{\ast}$ and Marcela Svarc$^{\ast\ast}${\footnote{Corresponding author:
Marcela Svarc, Departamento de Matem\'atica y Ciencias, Universidad de San Andr\'{e}s.... Email: msvarc@udesa.edu.ar}}}

\noindent\emph{$^{\ast}$Departamento de Matem\'aticas, Universidad
Aut\'{o}noma de Madrid, Spain}
\\[0pt]\emph{$^{\ast\ast}$Departamento de Matem\'atica y Ciencias,
Universidad de San Andr\'es and CONICET, Argentina}

\end{center}

\begin{abstract}
This paper presents SeqClusFD, a top-down sequential  clustering method for functional data. The clustering algorithm extracts the splitting information either from trajectories, first  or second derivatives. Initial partition is based on gap statistic that provides local information to identify the instant with more clustering evidence in trajectories or derivatives. Then functional boxplots allow reconsidering overall allocation and each observation is finally assigned to the cluster where it spends most of the time within whiskers. These local and global searches are  repeated recursively until there is no evidence of clustering at any time on trajectories or first and second derivatives. SeqClusFD simultaneously estimates the number of groups and provides data allocation. It also provides valuable information about the most important features that determine cluster structure. Computational aspects have been analyzed and the new method is tested on synthetic and real data sets.

\end{abstract}



\section{Introduction}\label{sec:introduction}

Functional data analysis (FDA) is a very active area of research nowadays, mainly since it has become very easy to collect and store data in ``continuous time'' (see \cite{C14}). Although generally each data is recorded only in a finite number of moments, it is more common to analyze then as functions rather than as vectors. Instead of points, the observations in this context are each of the curves. Most existing procedures for functional data require a large amount of consecutive observations on which smoothing techniques are applied, see \cite{RS05}. The classical assumption is that functions belong to a Hilbert space (for example, square integrable functions on a finite real interval $[a,b]$) and can be represented with a convenient functional basis. Depending on the characteristics of the functions, the most common basis are Fourier, Haar or wavelets. A finite truncation of the series facilitates the analysis with conventional multivariate methods applied to a finite set of retained coefficients.

The aim of cluster analysis, or unsupervised classification, is to assign observations into subsets, such that similar objects are in the same group and dissimilar ones are in different groups. This is a complex problem since usually there is no previous information about the data structure. There is no ``one size fits all'' method for analyzing any data set and the nature of the data should determine the procedure to be used. In fact, there are clustering methods that have outstanding performance in certain configurations of data and a terrible behavior under different conditions. For instance, it is well known that the popular $k$-means is suitable in $\mathbb{R}^d$ for round, Gaussian and well separated clusters but it is not able to find cluster structures with nested groups. Some multivariate clustering techniques have been successfully adapted to functional data after considering specific geometrical notions and can be classified in one of the next three categories.

\emph{Centroid based clustering methods} use optimization algorithms for centroid finding of each group (as $k$-means). Several authors studied and adapted these methods to functional data, see for instance \cite{ACMM93}, \cite{TK03}, \cite{CL07}, \cite{CF07}, \cite{TYI08}, \cite{GF10}, \cite{SSVV10a}, \cite{SSVV10b} and \cite{IPPV13}.

\emph{Model based clustering methods} assume different population model for each group, typically all from the same distribution family (as Gaussian mixture models). In FDA there are several procedures following this strategy, such as those proposed in \cite{JS03},\cite{BCH08} and \cite{JP14A}.

\emph{Hierarchical clustering methods} are based on the idea of divisive and/or agglomerative sequential grouping to provide the best partition on each possible number of groups. Divisive methods begin allocating all observations into one group and sequentially separate into different groups the observations that are more distant from the rest. This idea is repeated until there are as many groups as observations. Agglomerative methods initially assign each observation to a different group and then sequentially join the closest observations until all observations are in a single group. For functional data Giraldo \emph{et al.} \cite{GDM12} developed an algorithm for spatially correlated data and Boull\`e \emph{et al.} \cite{BGR14} introduced a bayesian proposal of nonparametric hierarchical clustering for functional data. At the best of our knowledge, the particular case of hierarchical clustering based on decision trees with sequential binary partitions,   have never been extended to FDA from multivariate data approaches where several extensions can be found, see for instance \cite{LXY00},\cite{BK05} and \cite{FGS13}.

In general, in centroid or model based methods the number of clusters is known while in hierarchical clustering is unknown. Recently, Jacques and Preda \cite{JP14B} described thoroughly the state of the art of this field.

\begin{figure}[!h]
\centering
\includegraphics[width=2.5in]{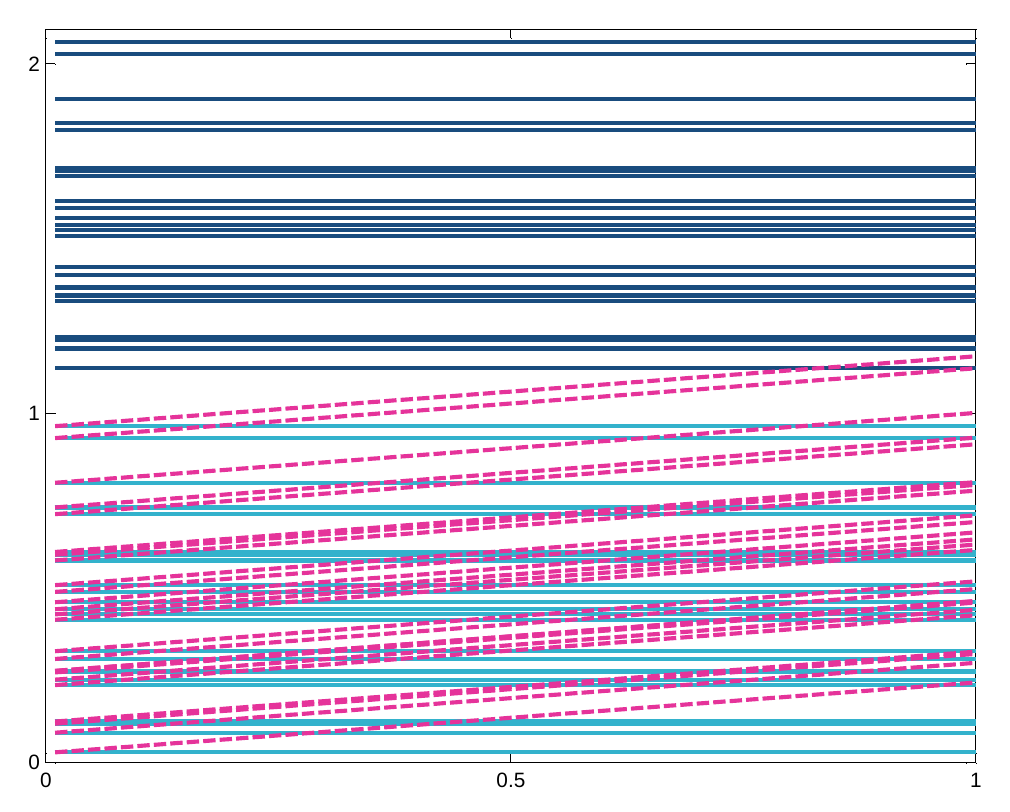}
\caption{Light solid lines are in group 1; dark solid lines are in group 2; and dashed lines are in group 3.} \label{toy}
\end{figure}

As it is well known, not all relevant information on a function is visible in the trajectory. Derivatives are usually very helpful to highlight differences. Although all clustering methods for functional data mentioned above can be applied to the set of trajectories, or the set of first derivatives, second derivatives, and so on, it is interesting to have a statistical procedure able to extract relevant information from any or all of these sets. In supervised classification of functional data, Alonso et al. \cite{ACR12} considered a distance base on using derivatives.  Exploration of functions and derivatives is a key point since group structure must be associated to the trajectories for some groups and to derivatives in some others. To illustrate this idea, lets analyze a very simple example with the three groups of 25 functions shown in Figure~\ref{toy}. Functions in groups 1 and 2 are constant but with different levels, while functions in group 3 have constant positive slope and levels are similar to functions in group 1. Considering original functions, only two groups are identified by looking in any instant, a cluster that contains lines of groups 1 and 3 and the other with lines from group 2. Switching to apply clustering methods with first derivative functions, we note that all lines in groups 1 and 2 have zero constant derivative function, different than derivatives from group 3 that are also constant but nonzero. Any clustering method applied to derivatives also identify a maximum of two clusters, one with the lines of groups 1 and 2 and the other with lines from group 3. In summary, by applying clustering methods to trajectories, functions are separated by level and when applied to derivatives are separated by shape. We will only be able to succeed in identifying the three groups if we run a sequential clustering scheme taking into account trajectories and derivatives.

The aim of this paper is to introduce \textbf{SeqClusFD}, a new hierarchical clustering method designed only for FDA.  The idea is to develop a divisive top-down clustering based on sequential exploration of the functions and their derivatives. This means that we consider local level and global shape properties that are available in functions, and not in finite dimensional data. Output from the algorithm includes number of groups, allocations and some guidelines for cluster interpretation.

The remainder of this paper is organized as follows. In Section 2, SeqClusFD method is presented and some details on practical implementation are analyzed. In Section 3, a simulation study with synthetic data is performed. In Section 4, well known real data are analyzed and a further study is conducted, including result interpretation. Conclusions are given in last Section.

\section{Divisive top-down SeqClusFD method}



We observe $X_1(t),\dots,X_n(t)$ functions grouped into an unknown number $k$ of populations, $C_1,\dots,C_k$.
We assume all the functions are smooth and defined in the same compact real interval $[a,b]$. Each data $X(t)$ is a realization of a stochastic process defined on the probability space $(\Omega,\emph{F},P)$,
hence $X(t)$ is a random variable for each  $t \in [a,b]$.

Two facts are motivating the need of a new sequential procedure for cluster identification that takes into account the shape of the curves. The first is that functions that belong to different groups must differ by at least one of the following characteristics: trajectories, velocities or accelerations. The second is that it may happen that some groups differ by the shapes of the trajectories and some others by the shapes on any of their derivatives. We use the notation  $X_1^l(t),\dots,X_n^l(t)$, where $l=0,1,2$ means the order of derivation. For the sake of simplicity, if $l=0$ we omit the superscript.

Hence, our idea is to start with a single group and, on each splitting stage, sequentially apply the following local and global clustering steps. We stop the divisions when there is no evidence of new cluster structure inside any group.

\noindent \emph{Local clustering step:}

Considering the sets $\{X_1(t), \dots,X_n(t)\}$, $\{X^1_1(t),\dots,X^1_n(t)\}$ and $\{X^2_1(t), \dots,X^2_n(t)\}$ separately, we find the instant $t\in T$ in which the cluster structure is more evident in any of the three sets. Local clustering evidence is assessed by computing for each set and instant the gap statistic (GS) introduced by Tibshirani \emph{et al.} \cite{TWH01} to estimate the number of clusters in a data set with any cluster method. The idea is to compare the change in the within cluster dispersion when one additional group is considered with the expected change under an appropriate reference distribution.


In our case $X^l_1(t),\dots,X^l_n(t)$ is a sample of one dimensional observations and for a $k$-cluster structure $C_1,\dots,C_k$, the within cluster dispersions are estimated by
$$
D_r=\sum_{i,i'\in C_r}d_{ii'}, \;\; r=1,\dots, k
$$
where $d_{ii'}$ is the euclidean distance between two observations of the same cluster.
The gain of considering $k+1$ groups instead of $k$ is
\begin{equation}
L_{k,k+1}=\ln (W_{k})-\ln (W_{k+1}),
\label{estadistico}
\end{equation}
where $W_k=\sum_{r=1}^k D_r/2n_r$ is the total sum of pairwise distances within $C_1,\dots,C_k$ and $n_r$ is the cardinal of $C_r$.

%

As large $L_{k,k+1}$ indicates evidence of $k+1$ groups, this value is compared with that obtained from a sample generated from an appropriate reference distribution. Tibshirani \emph{et al.} \cite{TWH01} proved that, for the one-dimensional case, the reference distribution should be uniform. Hence we generate $b=1,\dots,B$  random samples of size $n$ from the uniform distribution on the interval $[\min X_i,\max X_i]$. For each sample we compute $\ln (W_{k,b})$ and $\ln (W_{k+1,b})$. We denote the empirical version of (\ref{estadistico}) by  $L^*(k,k+1)=E^*(\ln (W_{k,b}))-E^*(\ln (W_{k+1,b}))$, where
$E^*(\ln (W_{kb}))$ and $E^*(\ln (W_{k+1,b}))$ are empirical means. Empirical standard deviation of $\ln (W_{k,b})$ is denoted by  $s^*(k)$.

The gap statistic for the set $X^l_1(t),\dots,X^l_n(t)$ is the minimum $k$ that satisfies
\begin{equation}
L^*(k,k+1)+n_{sd} \sqrt{1+1/B}s^*(k+1)\geq L(k,k+1),
\label{GS}
\end{equation}
where $n_{sd}$ is the number of standard deviations to be fixed. In \cite{TWH01}, the authors  suggest  $n_{sd}=1$, but stronger rules could be considered.

We use the greatest gap statistic to decided the instant, feature (trajectory, first or second derivative) and number of groups for sample splitting. Then for the local tentatively classification we apply the one-dimensional $k$-means allocation on these instant to all the curves of this feature.

\noindent \emph{Global revision clustering step:}

The previous functional data classification is only based on local information, then we need a global criteria to integrate the information from all the complete curves and correct possible local effects.

Reallocation of misleading data is done by identifying potential outliers using the functional boxplot (FB) introduced by Sun and Genton \cite{SG11}. For each set of the proper feature curves inside the same cluster identified in the local clustering step we compute a FB. Construction is based on ordering functions from the center outward using band-depth definition of L\'{o}pez-Pintado and Romo \cite{LP09}. The appearance is determined by deepest curve (\emph{median}), envelope of $50\%$ central functions (\emph{box borders}) and maximum non-outlying envelope (\emph{whiskers}). In analogy with classical boxplots, 1.5 times the $50\%$ central region is typically considered the whiskers for the non-outlying envelope. In \cite{LP09} the authors suggest that this value can be adjusted on practice.

We consider potential outlier to any curve that is outside the whiskers band at least for one instant. For these curves we compute the proportion of time that are inside each FB. In the final cluster allocation, they are moved to the group where they spend most of the time inside whiskers.

\subsection{SeqClusFD algorithm}





\noindent STEP 0.
\begin{enumerate}
\item[-] \emph{Define} a grid on the interval $[a,b]$: $a=t_0 < \dots < t_N=b$.

\item[-] \emph{Evaluate} trajectories, first and second derivatives in the grid: $X_i^l (t_j)$ for $i=1,\dots,n$, $l=0,1,2$ and $j=0,\dots,N-l$.

\item[-] \emph{Start} by considering a single group.
\end{enumerate}

\

\noindent STEP 1. {\emph{Local clustering.}}
\begin{enumerate}
\item[-] \emph{Calculate} gap statistics $\widetilde{k}(A_{j,l})$ in combination with $k$-means clustering for all possible data sets $A_{j,l}=\{ X_1^l (t_j),\dots, X_n^l (t_j)\}$,  where $l=0,1,2$ and $j=0,\dots,N-l$, .

\item[-] \emph{Estimate} the number of groups in the functional data set by $\hat k = \max_{j,l} \widetilde{k}(A_{j,l})$.

\item[-] \emph{Set up} $\hat C_1,\dots,\hat C_{\hat k}$  with the complete proper feature functions associated to the $k$-means clusters of $ A_{\hat j,\hat l} = arg \max_{j,l} \widetilde{k}(A_{j,l})$. In case of ties, choose $ A_{\hat j,\hat l}$ that provides more evidence of $\hat k$ clusters, this means that has larger value of $L_{\hat k, \hat k+1}(j,l)$.

\end{enumerate}


\noindent STEP 2. {\emph{Global cluster revision.}}
\begin{enumerate}
\item[-] \emph{Compute} $\hat k$ functional boxplots with the curves inside $\hat C_1,\dots,\hat C_{\hat k}$. Consider 3 times point wise interquartil range for maximum length of whiskers, which is the usual rule in one dimensional boxplot to identify extreme outliers. For $r=1,\dots,\hat{k}$, call $(LB_r(t),UB_r(t))$ to the lower and upper extreme outlier whisker bands of functional boxplots.

 \item[-] \emph{Identify} potential outliers as functions outside $(LB_r(t),$ $UB_r(t))$, at least for one instant, for $r=1,\dots,\hat{k}$.

 \item[-] \emph{Reallocate} each potential outlier into the cluster $\hat C_1,\dots,\hat C_{\hat k}$ where spends more time inside $(LB_r(t),UB_r(t))$, for $r=1,\dots,\hat{k}$.
\end{enumerate}


\noindent STEP 3.

If one group is divided, repeat step 1 and 2 for this group to find possible partitions. Stop division when gap statistic in step 1 determines that there is only one group for every instant $t_0< \dots < t_N$ and feature of the data set (trajectories or derivatives).

\subsection{Practical implementation}

Considering again the example shown in Figure \ref{toy}, we find that SeqClusFD separates in the first step the third group using derivative functions. In second step, SeqClusFD separates the first two groups considering the trajectories. In addition, SeqClusFD determines that there are no further groups. Then, three groups are found and there are not misclassified observations. Ieva et al \cite{IPPV13}, perform a $k$ means clustering procedure for functional data, but instead of using the classical $L^2$ norm they work in the Sobolev $H^1$ space with the ordinary norm, this means that they consider simultaneously the information of the function and the derivative. If we use that approach in this example, and estimate the number of clusters either by the Gap Statistics or by Calinski-Harabasz approach, which are among the best well known procedures to estimate the number of clusters, neither of them detect the correct number of clusters.

In more realistic problems computation of the first two derivatives is not a simple task. Typically, functional data are characterized by a high amount of consecutive observations defined in an interval of finite length, but frequency may defer from one individual to another and  observations are not necessarily equidistant. When measures do not have sampling errors, function values can be interpolated and then derivatives are computed by differentiating. Otherwise it is necessary to use smoothing techniques in order to transform observations into functions that can be evaluated for any time $t$. Different smoothing techniques should be considered since each data set has different properties, as for instance functions could be monotone, or periodic or non-negative. Ramsay \emph{et al.}, \cite{RHG09} discuss the main options that are consider in FDA, most of them implemented in Matlab and R routines that can be downloaded from \url{http://www.functionaldata.org}.


For every $t_j$ and feature of the trajectories, uniform distribution boundaries for the bootstrap samples, that are necessary for gap statistic computations, will probably be different. Three bootstrap procedures should be executed for each $t_j$. Then the thinner the grid $t_0 < \dots < t_N$ defined in step 0 of SeqClusFD algorithm more computationally expensive is step 1. To speed up SeqClusFD we suggest pointwise rescaling to the interval $[0,1]$ for observations in subsets $A_{j,l}=\{ X_1^l (t_j),\dots, X_n^l (t_j)\}$ ($l=0,1,2$ and $j=0,\dots,N-l$) as follows,
$$
Y_i^l(t_j)=\frac{X_i^l(t_j)-\min \{A_{j,l}\}}{\max \{A_{j,l}\}-\min \{A_{j,l}\}}, \mbox{ for } i=1,\dots,n.
$$
Now all bootstrap samples should be generated according to a uniform distribution on the interval $[0,1]$. This means that  bootstrap procedure can be done only once for each visit of the algorithm to step 1.  However, when step 1 is repeated it is because a group is divided and then sample size is different. We can not avoid bootstrap should be redone on different visits.

\section{Simulation Study}

To show the performance of SeqClusFD we conducted a simulation study using different artificial data sets that have been previously proposed in the literature for clustering of functional data. In all cases we report the number of times that SeqClusFD selects the correct number of groups. For these successful examples, the correct classification rate (CCR) is estimated.

As SeqClusFD is based on gap statistics and functional boxplot, before executing the algorithm it is necessary to fix the value of some parameters related to these tools. For gap statistic we follow suggestions in the original paper, except for $n_{sd}=3$ on equation (\ref{GS}) since we are applying a stricter rule for identifying clusters with strong evidence at a certain time period. More that 500 bootstrap samples are not necessary. For functional boxplot we follow the classical rule to identify severe outliers, considering the maximum length of the whiskers as three times the interquantil range.

To reduce computational effort in the local clustering step we limit the maximum number of clusters to five. It is important to remark that, even though this parameter fixes an upper bound for the number of clusters on each partitioning step, there is not an upper bound for the general procedure. In the global cluster revision step, A minimum of 10 functions are required to compute a boxplot. The minimum cluster size to continue with splitting is also fixed on 10 functions

We propose the same values for all the simulations and real data examples.

\subsection{Sampling errors associated with the entire curve}

Sangalli \emph{et al.} \cite{SSVV10a}  introduced three models for curve generation. We follow their models for data generation  with the idea of exploring the SeqClusFD performance in clusters structures with different amplitudes and curve registration problems.

\begin{figure}[!t]
\centering
\includegraphics[width=5.5in]{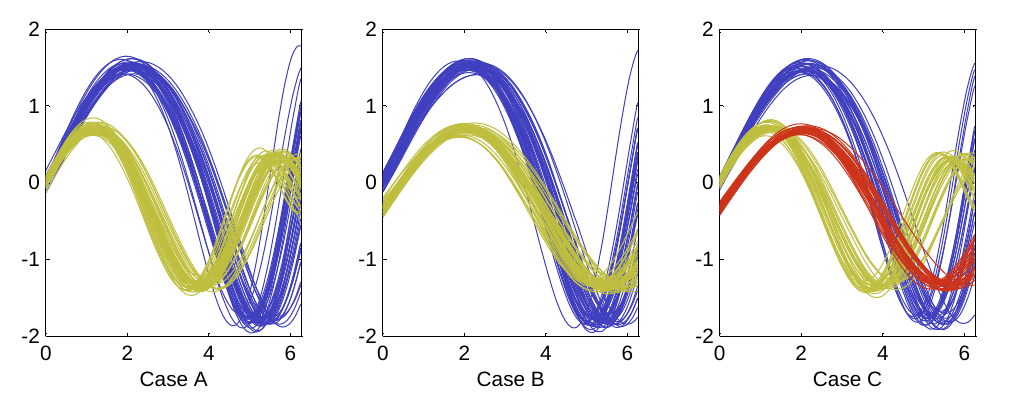}
\caption{Examples of data set of functions simulated with models A, B and C.} \label{curvassangalli}
\end{figure}


\noindent \emph{Model A:} Two clusters with $n/2$ functions generated as follows,
\begin{eqnarray}
 X_{i}(t)&=&(1+\epsilon_{1i})\sin\left(\epsilon_{3i}+ \epsilon_{4i}t\right) + \nonumber \\
         && (1+\epsilon_{2i})\sin\left(\frac{\left(\epsilon_{3i}+\epsilon_{4i}t\right)^2}{2\pi}\right) \label{grupoA} \\
 &&  \mbox{ for } i=1,\dots,n/2,  \nonumber  \\
    X_{i}(t)&=&(1+\epsilon_{1i})\sin\left(\epsilon_{3i}+\epsilon_{4i}t\right)- \nonumber \\
    && (1+\epsilon_{2i})\sin\left(\frac{\left(\epsilon_{3i}+\epsilon_{4i}t\right)^2}{2\pi}\right)
 \label{grupoB} \\
 && \mbox{ for } i=n/2+1,\dots,n. \nonumber
\end{eqnarray}

\noindent \emph{Model B:} Two clusters with $n/2$ functions generated in the first group following (\ref{grupoA}) and in the second group as follows,
\begin{eqnarray}
X_{i}(t)&=&\left(1+\epsilon_{1i}\right)
\sin\left(\epsilon_{3i}+\epsilon_{4i}\left(-\frac{1}{3}+\frac{3}{4}t\right)\right)- \nonumber \\
&& \left(1+\epsilon_{2i}\right) \sin\left(\frac{\left(\epsilon_{3i}+\epsilon_{4i}\left(-\frac{1}{3}+\frac{3}{4}t \right)\right)^2}{2\pi}\right) \label{grupoC} \\
&& \mbox{ for } i=n/2+1,\dots,n.   \nonumber
\end{eqnarray}

\noindent \emph{Model C:}  Three clusters with $n/3$ functions generated in the first group following (\ref{grupoA}), in the second group following (\ref{grupoB}) and in the third group following (\ref{grupoC}).

We simulate 200 data sets of size $n=90$ for each model.  All errors $\epsilon_{1i},\dots,\epsilon_{4i}$ are independent and normally distributed with mean 0 and standard deviation 0.05. Figure \ref{curvassangalli} shows three examples of 90  curves generated according with models A, B and C.

Table \ref{resultsimul1} reports for each model, percentage of data sets in which SeqClusFD finds the number of clusters indicated on first column. In bottom line are displayed the mean of correct classification rates (CCR) calculated only with data in which the number of clusters is correctly identified. In almost all cases it find the correct number of clusters. The result is poorer in the model C,  which is also the most challenging problem, with a 75.5\% of successful identifications. When the number of clusters is not identified correctly, generally it is determined that there is an additional cluster. We observed that in these cases one of the original cluster is divided into two and the other clusters are identified without error.  Table \ref{resultsimul1} also shows the mean of correct classification rates (CCR), calculated only with data sets in which the number of clusters is correctly identified. It is clear that SeqClusFD has an outstanding performance in these cases.

\begin{table}[!t]
\renewcommand{\arraystretch}{1.3}
\caption{Distribution of the Number of Groups found by SeqClusFD and Mean CCR.}
\label{resultsimul1}
\centering
\begin{tabular}{c||ccc}
\hline
\textbf{Number of clusters}  & \textbf{Model A} & \textbf{Model B} & \textbf{Model C} \\ \hline \hline
  2 & $88$ & $99.5$ & $0$ \\
  3 & $12$ & $0.05$ & $75.5$\\
  4 & $0$ & $0$ & $23.5$\\
  5 & $0$ & $0$ & $1$\\
  \hline  \hline
\textbf{Mean CCR} & $99.67$ &  $99.96$ & $99.45$ \\
 \hline
\end{tabular}
\end{table}

Finally, we compare our procedure with other clustering techniques available in R for functional data, namely, \textit{funclust} (\cite{JP14A}), \textit{funHDDC} (\cite{BJ11}) and \textit{kmeans.fd} which is a k--means procedure for functional included in the R package \textit{fda.usc} (\cite{FO12}). In all the cases the number of clusters must be given beforehand.  Table \ref{compsimul} exhibits the mean CCR for the three procedures proposed. It is clear that SeqClusFD outperforms the three clustering strategies since it has higher CCR for all the models considered.

\begin{table}[!t]
\renewcommand{\arraystretch}{1.3}
\caption{Mean CCR for funclust, kmeans.fd and funHDDC.}
\label{compsimul}
\centering
\begin{tabular}{c||ccc}
\hline
  & \textbf{Model A} & \textbf{Model B} & \textbf{Model C} \\ \hline \hline
  funclust & $75.62$ & $73.57$ & $48.48$ \\
  kmeans.fd & $67$ & $76.77$ & $68.45$\\
  funHDDC & $85.81$ & $47.39$ & $37.86$\\
  \hline
 \hline
\end{tabular}
\end{table}

%
%
%
%

\subsection{Sampling errors associated to each instant}

We simulate 200 data sets with four clusters that are generated with a similar model to that introduced by Serban and Wasserman \cite{SW05}. Each cluster contains 150 functions that are generated as follows
\begin{eqnarray}
X_{ij}(t)&=&f_j(t)+\epsilon_i(t), \nonumber \\
&& \mbox{ for } t \in [0,1],  i=1,\dots,150  \mbox{ and } j=1,\dots,4, \nonumber
\end{eqnarray}
where
$$
f_1(t)  =  min\left(\frac{2-5t}{2},\left(\frac{2-5t}{2}^2 \sin\left(\frac{5 \pi t}{2}\right)\right)\right),
$$
$$
f_2(t)  =  -f_1(t), \;\; f_3(t)  =  \cos(2\pi t) \;\; \mbox{ and } \;\; f_4(t) = -f_4(t).
$$


In \cite{SW05} the authors consider independent errors, while we consider correlated errors from a Gaussian process. For all the functions $\epsilon(t)$ has normal distribution with mean 0.4, standard deviation $0.9$ and covariance structure given by,
$$
\rho \left( s,t\right) =0.3\exp \left( -\frac{( s-t)^2 }{0.3}\right) ,\text{ \ \ for }s,t\in \left[ 0,1\right].
$$

\begin{figure}[!t]
\centering
\includegraphics[width=5.5in]{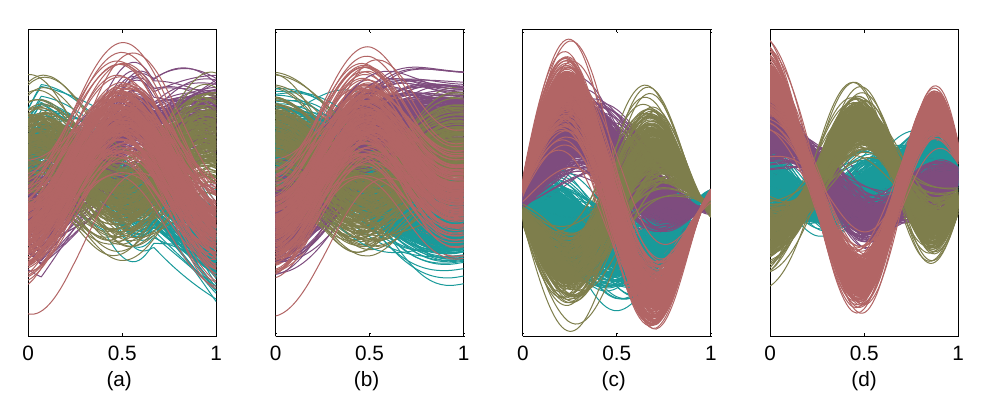}
\caption{(a) Simulated data set with sampling errors associated to each instant; (b) smoothed data set; (c) first derivatives and (d) second derivatives.} \label{grafwasserman}
\end{figure}

Figures \ref{grafwasserman}(a) and \ref{grafwasserman}(b) show one of the generated data set and the corresponding  smoothed data set with $B$-splines, respectively. Each color represents one theoretical cluster. In Figures \ref{grafwasserman}(c) and \ref{grafwasserman}(d) are displayed the firsts two derivatives. To prevent boundary effects we reflect one third of the observations at the beginning and at the end of each measurement.
We also challenged our method with other clustering procedures for functional data: \textit{funclust, funHDDC} and \textit{kmeans.fd}, in these case the number of clusters was given as an input. Table \ref{resultWasser} exhibits the mean CCR for the 200.  It is important to highlight that SeqClusFD always identifies four groups and that has a much more higher  mean correct classification rate  than the rest of the procedures, achieving  perfect classification in most of the replicates.

 \begin{table}[!t]
\renewcommand{\arraystretch}{1.3}
\caption{Mean CCR for funclust, kmeans.fd, funHDDC and SeqClusFD.}
\label{resultWasser}
\centering
\begin{tabular}{c||ccc}
\hline
  Clustering Procedure & Mean CCR \\ \hline \hline
  funclust & $ 40.05$ \\
  kmeans.fd & $60.12$ \\
  funHDDC & $63.21$ \\
  SeqClusFD & $99.85$ \\
  \hline
 \hline
\end{tabular}
\end{table}

\section{Real Data Examples}

\subsection{Berkeley Growth Study data}

\begin{figure*}[!t]
\centering
    \subfloat[Growth trajectories]
    {\includegraphics[width=2.5in]{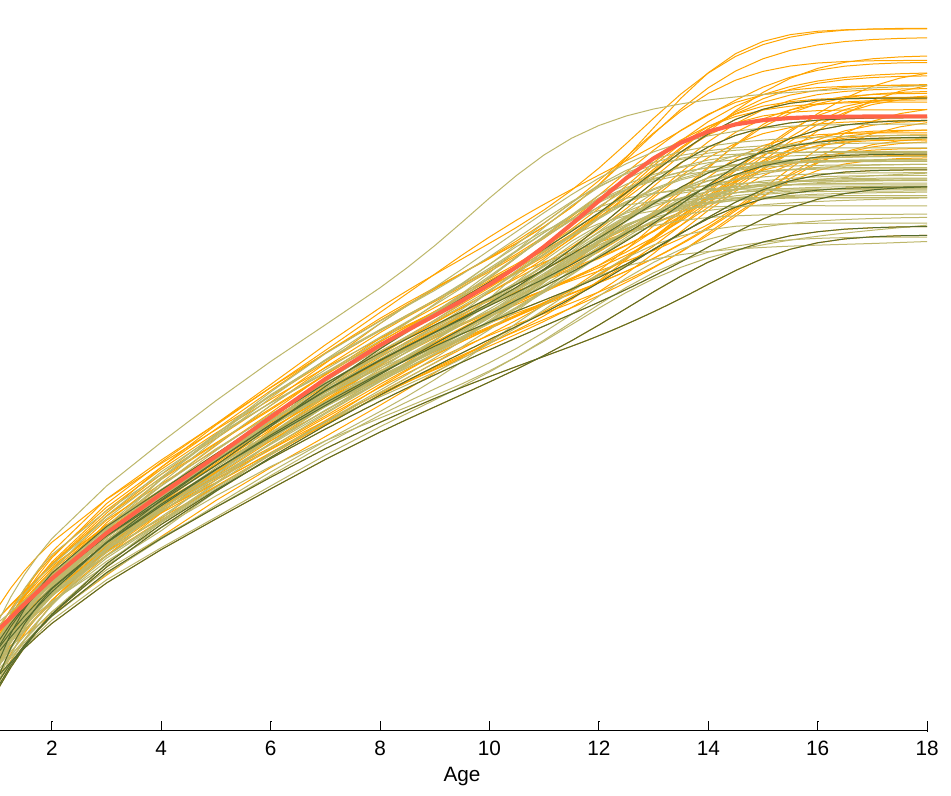}
    \label{growthtr}}
    \hfill
    \subfloat[Growth derivatives]{\includegraphics[width=2.5in]{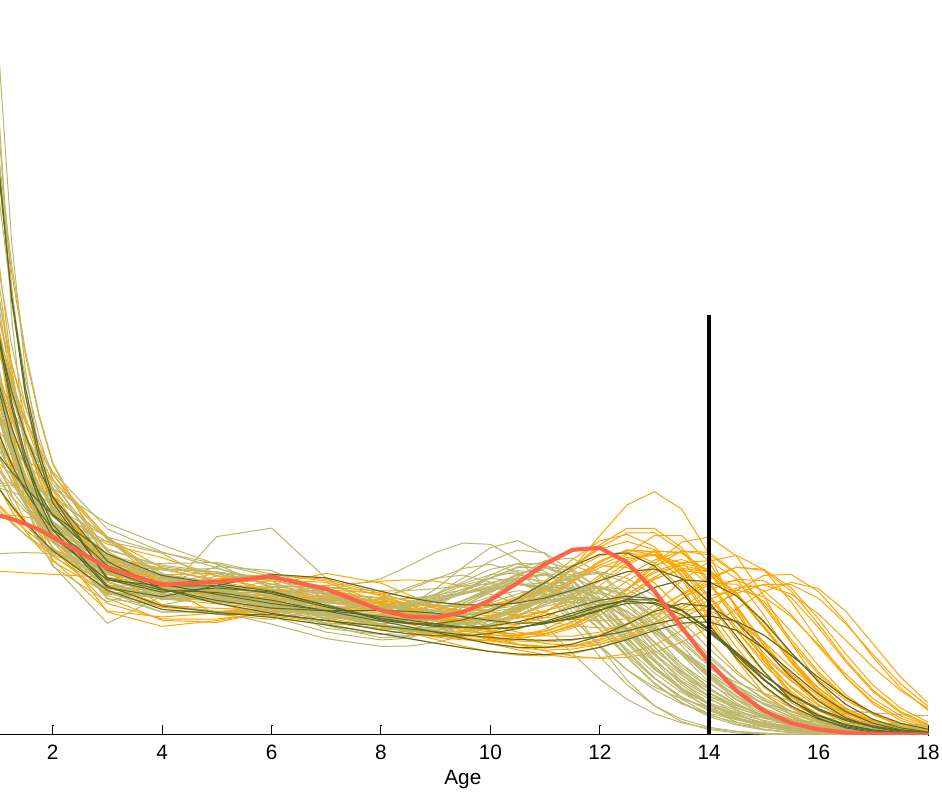}
    \label{growthderiv}}
    \label{growth}
    \caption{Heights of 54 girls (green curves) and 39 boys (orange curves) measured between 1 and 18 years. Misclassified curves by SeqClusFD are highlighted with thicker lines.}
  \end{figure*}

The Berkeley Growth Study is one of the best known long-term development research ever conducted, and the height growth data set introduced by Tuddenham and Snyder \cite{TS54} is a reference to illustrate different methods for FDA.
In particular, the heights of 54 girls and 39 boys measured between 1 and 18 years in 31 unequally spaced moments
(see Figure \ref{growthtr}) are considered one of the most challenging data sets for clustering purposes. More measurements were taken when growth was more rapid, as childhood and adolescence, and least in the early years, when growth was more stable. As a consequence of that, for the computation of the first two derivatives, we need to transform observations into functions that can be evaluated for any time. We consider a monotonic cubic regression spline smoothing as suggested by Ramsay \emph{et al.} \cite{RHG09}.


Our first objective is to test the effectiveness of SeqClusFD in determining the number of groups in this data set without regard the gender information. Second, we use the output of SeqClusFD for classifying children. The information from step 1 gives us the key for understanding how the groups differ.

Using the same parameters as in simulation study for step 1 and 2 of SeqClusFD, final output identifies two clusters. The algorithm finds in the local step the highest evidence of clustering in the first derivative (growth speed) at the age of 14, which is coincident with the end of puberty in girls but not in boys yet. In the final classification of the data, one cluster is coincident with boys and the other with girls. Only 10 curves are misclassified and are highlighted with thicker lines in Figures  \ref{growthtr} and \ref{growthderiv}. Misclassifications correspond to a boy with early puberty and 9 girls with late maturity. The correct classification rate is CCR $= 89.25\%$.

This data set has been used recently by Jacques and Preda \cite{JP14A} to compare several clustering methods for functional data. All analyzed methods make use of the information that the number of clusters is two, which could suggest that \emph{a priori} SeqClusFD is less competitive. However the CCR for alternative methods are not always superior:  funclust-CCR $ =69.89\%$ (Jacques and Preda, \cite{JP14A});  FunHDDC-CCR $ = 96.77\%$ (Bouveyon and Jacques, \cite{BJ11}); fclust-CCR $= 69.89\%$ (James and Sugar, \cite{JS03}); and kCFC-CCR $=93.55\%$ (Chiou and Li, \cite{CL07}). SeqClusFD is closer to the most successful (FunHDDC  and KCFC) than to funclust and fclust. In addition, the output of SeqClusFD also includes an estimate of the number of groups and provides assistance for the interpretation of the clusters.

Sangalli \emph{et al.} \cite{SSVV10b} also analyzed this data set. Although they find more evidence for the existence of a single cluster, they analyze the case of two clusters and the best CCR obtained is $88.17\%$.

\subsection{EGC200 data}

\begin{figure}[!t]
\centering
\includegraphics[width=5.5in]{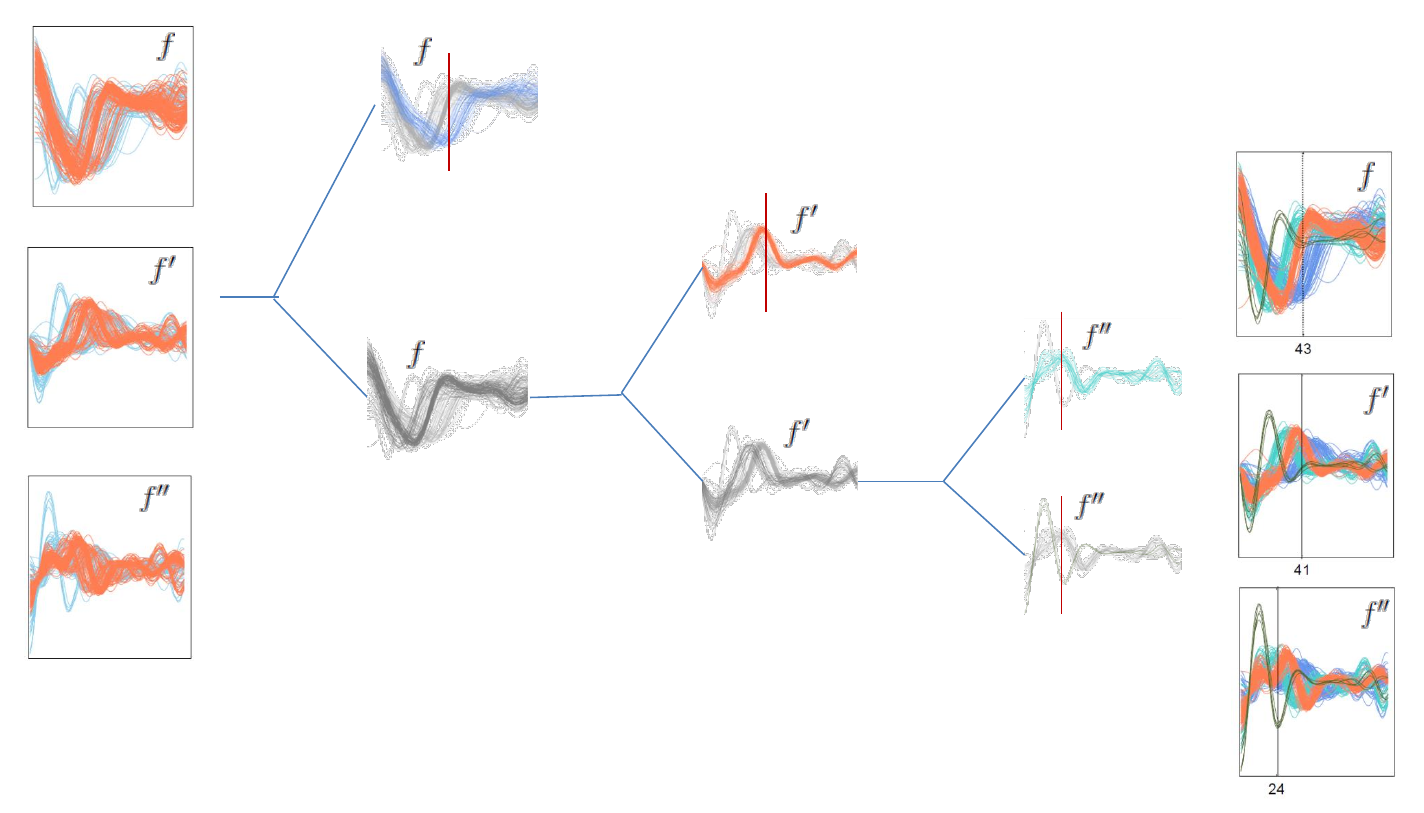}
\caption{On the left side, electrocardiograms ($f$), first ($f'$) and second ($f''$) derivatives for the two clusters (blue and orange) of ECG200 data set. On the central part, results of the three iterations executed by SeqClusFD. On the right side, final classification with SeqClusFD.} \label{grafeggteoemp2}
\end{figure}

The 200 electrocardiograms  of ECG200 data set can be found in the UCR Time Series Classification and Clustering website \cite{CKHetal15}. Data set has two groups, with 133 and 67 electrocardiograms  each one, all of them recorded at 96 equally spaced instants. Left side of Figure \ref{grafeggteoemp2} shows electrocardiograms ($f$) and their first ($f'$) and second ($f''$) derivatives for the two clusters (orange and blue curves).  These data has been analyzed by Jacques and Preda \cite{JP14A}, among others, using the same clustering procedures than in the previous example, except kCFC. The results are: funclust-CCR $= 84\%$; FunHDDC-CCR $=75\%$; and fclust-CCR $=74.5\%$.

Tree structure in central part of Figure \ref{grafeggteoemp2} shows the results on the three iterations needed by SeqClusFD to complete the clustering procedure. We use the same parameters as in simulation study for step 1 and 2. In the first iteration, SeqClusFD algorithm detects two groups using the information provided by the electrocardiograms at $t=43$. When the algorithm visits again step 1 for the first group, gap statistic determines that there is only one cluster and SeqClusFD stops division. The majority of the curves in this cluster are blue, 35, and 19 curves are orange. In the second iteration, the other group is divided in two clusters at $t=41$ with the information of the first derivatives. One cluster is a terminal node with 99 orange curves and only 7 blue that could be considered as misclassifications. In the third iteration, the remaining observations are separated in two groups at $t=24$ considering the information of the second derivatives. The green curves are a terminal cluster with 21 blue curves and 15 oranges. Finally, last four curves can not be separated because of the small size of the group (under 10).  When exploring which are these curves we find that they can be considered as outliers. Some authors, see for instance \cite{JP14A}, eliminate these electrocardiograms from the beginning of the analysis.

Right side of Figure \ref{grafeggteoemp2} shows the final three cluster allocation in $f$, $f'$ and $f''$. The four outliers appear in dark green. Even though this data set only contains two clusters and we have found three, seen in the results that the group found in second iteration can be considered the same as the original cluster of orange curves,  with 7 curves that are misclassified.  Assuming that the original cluster of blue electrocardiograms is the union of the other two (blue and green), the CCR is $ 79.08 \% $. For this calculation we have excluded the four outliers in order to compare SeqClusFD results with those of Jacques and Preda \cite{JP14A}.

The result of SeqClusFD classification is very good in comparison with the other methods. Although SeqClusFD--CCR is not the highest CCR, note that the other methods start from an advantageous position by assuming the number of clusters is known. SeqClusFD is only overcome by funclust.

\subsection{Italy Power Demand  data}

\begin{figure}[!t]
\centering
\includegraphics[width=5.5in]{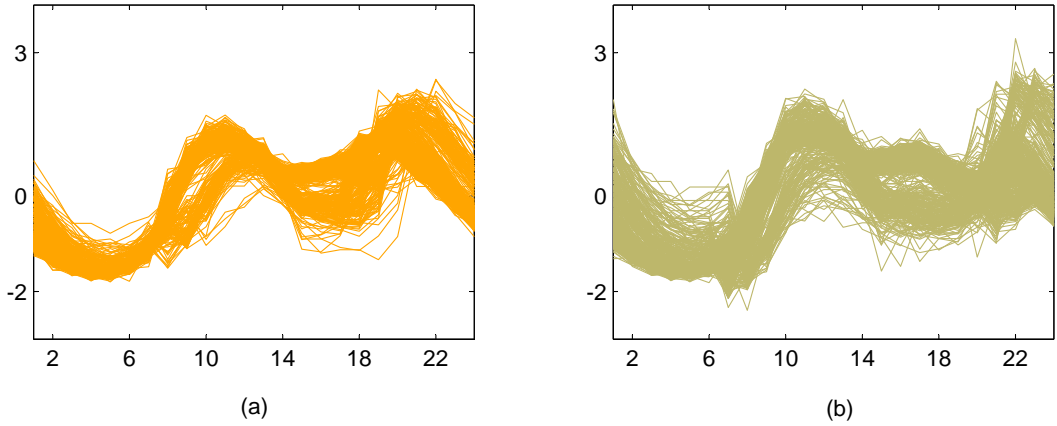}
\caption{Power demand curves along a day in Italy: (a) 513 curves in cluster 1; and (b) 516 curves in cluster 2.}
\label{italiaorig}
\end{figure}

Italy Power Demand data set can also be found in the UCR Time Series Classification and Clustering website \cite{CKHetal15}. Data set contains two clusters: cluster 1 with 513 curves of power demand (see Figure \ref{italiaorig}(a)); and cluster 2 with 516 curves (see Figure \ref{italiaorig}(b)). Note that inside each cluster there are two different pattern of consumers. All curves were measured at 24 equally spaced moments along a day. To increase the grid size on step 0 of SeqClusFD algorithm, data are smoothed with six equally spaced knot splines (choosing more knots lead to similar results). To prevent boundary effects we reflect one third of the records at the beginning and end of each record. SeqClusFD is executed with the same parameters as in previous examples.

\begin{figure}[!t]
\centering
\includegraphics[width=5.5in]{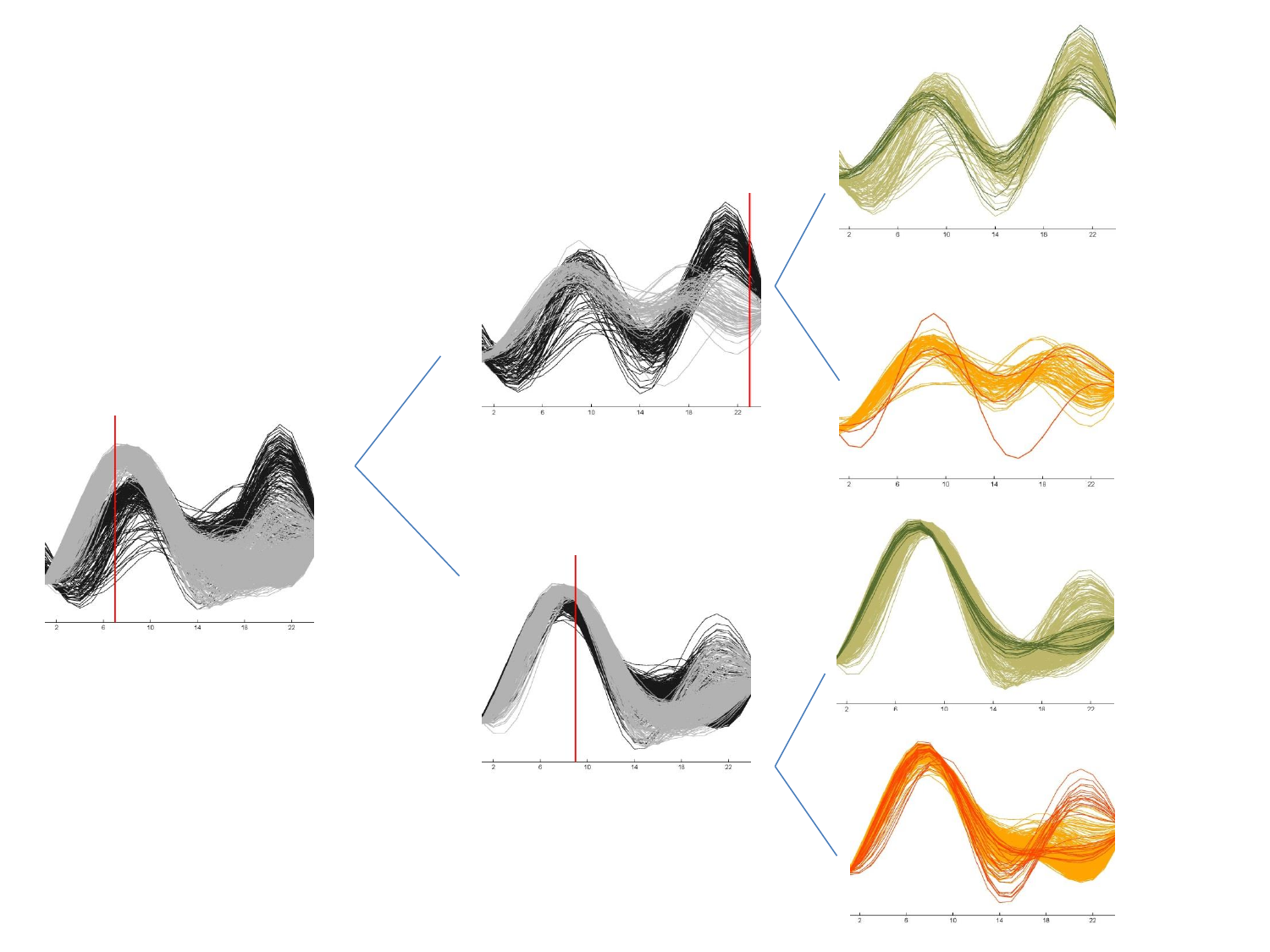}
\caption{Results of the two iterations executed by SeqClusFD to classify curves of power demand in Italy. First derivatives are shown in all iterations.}
\label{italiatree}
\end{figure}

SeqclusFD finds 4 clusters in the two iterations shown in the tree of Figure \ref{italiatree}, identifying the original clusters and the two different pattern of consumers within them. The partitions are always based on information provided by first derivatives, at $t=7$ in the first iteration and  $t=23$ and $t=9$ in the second. The final classification is shown on the right side of Figure \ref{italiatree}. Orange curves are cluster 1, while green curves are cluster 2. The darker green and orange curves are misclassified power demand curves. Collapsing the four clusters in only two we can calculate CCR, which is $93.49 \%$.

%
%

\section{Conclusion}

We introduce SeqClusFD, a hierarchical clustering algorithm for functional data that simultaneously determines the number of groups and the clustering conformation. As we are interested in cluster structures that takes into account the shape of the curves, we also consider the speed and acceleration functions linked to the original curves. It is a sequential procedure, that successively runs two steps, it searches the instant of time, along the functions and its derivatives, with most clustering ability, according to the gap statistics criterion and one dimensional k-means algorithm. Once this initial groups are conformed it builds up a functional boxplot for each group, with the objective of reallocating possibly misclassified data. Moreover, it provides helpful information towards the  grouping structure.


Although the SeqClusFD is based on the one dimensional gap statistic, alternative methods to estimate the number of groups could be considered. Similarly, the functional boxplot could be adapted to consider different depth definitions for functional data. Several proposals can be found in \cite{M13}.

The algorithm could be easily extended to the multivariate case, where each data is a vector of functions (see \cite{BJS11}). In the local step, any clustering method could be considered. On the global revision step, the boxplot could be designed using band depth proposed  by Lopez-Pintado \emph{et al.} \cite{LSLG14} for multivariate functional data. The efficiency of the method should decrease with the dimensionality.

Finally, the output of the algorithm exhibits number of groups and clustering allocation. In addition, it gives information about the \textit{key} moments $t\in [a,b]$ and features, i.e. $X(t)$, $X^1(t)$ or $X^2(t)$. This yields to a better understanding of the clustering structure. This topic is closely related to the variable selection problem that has been widely studied recently in the supervised classification framework, see for instance Tian and James \cite{TJ13} and Martin-Barragan \emph{et al.} \cite{MLR14}. However, this problem has not been analyzed for unsupervised classification problems, at the best of our knowledge.


\section*{Acknowledgments}
Ana Justel is supported by MINECO (Spain), grants CTM2011-28736/ANT and CTM2013-47381-P


\begin{thebibliography}{1}


\bibitem{ACMM93} C.~Abraham, P.A.~Cornillon, E.~Matzner-L{\o}ber and N.~Molinari, ``Unsupervised curves clustering using B-Splines'', \textit{Scandinavian Journal of Statistics}, 30, 581--595, 1993. DOI: 10.1111/1467-9469.00350


\bibitem{ACR12} A.M.~Alonso, D.~Casado and J.~Romo, ``Supervised classification for functional data: a weighted distance aprproach'', \textit{Computational Statistics and Data Analysis}, 56, 2334--2346, 2012. DOI: 10.1016/j.csda.2012.01.013

\bibitem{BK05} J.~Basak and R.~Krishnapuram, ``Interpretable hierarchical clustering by construction an unsupervised
decision tree''. \textit{IEEE Transactions on Knowledge and Data Engineering}, 17, 121--132, 2005. DOI:10.1109/TKDE.2005.11

\bibitem{BJS11} J.R.~Berrendero, A.~Justel, and M.~Svarc, ``Principal components for multivariate functional data''.  \textit{Computational Statistics and Data Analysis}, 55, 2619--2634, 2011. DOI:10.1016/j.csda.2011.03.011


\bibitem{BCH08} J.G.~Booth, G.~Casella and J.P.~Hobert, ``Clustering using objective functions and stochastic search'', \textit{Journal of the Royal Statistical Society, B}, 70, 119--139, 2008. DOI: 10.1111/j.1467-9868.2007.00629.x.

\bibitem{BGR14} M.~Boull\'e, R.~Guigour\`es,  and  F.~Rossi, ``Non parametric hierarchical clustering of functional data''. \textit{Advance in Knowledge Discovery and Management, Studies in Computational Intelligence}, 5, 15--35, 2014. DOI:10.1007/978-3-319-02999-32.

\bibitem{BJ11} C.~Bouveyron and J.~Jacques,  ``Model-based clustering of time series in group-specific functional subspaces''. \textit{Advances in Data Analysis and Classification}, 5, 281--300, 2011. DOI:10.1007/s11634-011-0095-6.

\bibitem{CKHetal15} Y.~Chen, E.~ Keogh, B.~ Hu, N.~Begum, A.~ Bagnall, A.~ Mueen and G.~ Batista. The UCR Time Series Classification Archive, 2015. (\url{http://www.cs.ucr.edu/~eamonn/time_series_data/})

\bibitem{CL07} J.M.~Chiou, and P.L.~Li, ``Functional clustering and identifying substructures of longitudinal
data''. \textit{Journal of the Royal Statistical Society, B}, 69, 679--699, 2011.  DOI: 10.1111/j.1467-9868.2007.00605.x.

\bibitem{CF07} J.A.~Cuesta Albertos and R.~Fraiman,  ``Impartial Trimmed $k$-means for Functional Data''. \textit{Computational Statistics and Data Analysis}, 51, 4864--4877, 2007. DOI:10.1016/j.csda.2006.07.011

\bibitem{C14} A.~Cuevas, ``A partial overview of the theory of statistics with functional data''. \textit{Journal of Statistical Planning and Inference}, 147, 1-23, 2014. DOI: 10.1016/j.jspi.2013.04.002.



\bibitem{FGS13} R.~Fraiman, B.~Ghattas and M.~Svarc, ``Interpretable clustering using unsupervised binary trees''. \textit{Advances in Data Analysis and Classification}, 7, 125--145, 2013. DOI:10.1007/s11634-013-0129-3

\bibitem{FO12} M.~Febrero-Bande, M. Oviedo de la Fuente, ``Statistical Computing in Functional Data Analysis: The R Package fda.usc''. \textit{Journal of Statistical Software},51(4), 1-28 ,2012.  DOI: 10.18637/jss.v051.i04


\bibitem{GF10} C.~Genolini and B.~Falisard, ``KmL: k-means for longitudinal data''. \textit{Computational Statistics}, 25, 317--328, 2010- DOI : 10.1007/s00180-009-0178-4.

\bibitem{GDM12} R.~Giraldo, P.~Delicado and J.~Mateu, ``Hierarchical clustering of spatially correlated functional data''. \textit{Statistica Neerlandica}, 66, 403--421, 2012. DOI: 10.1111/j.1467-9574.2012.00522.x.

    \bibitem{IPPV13} F. Ieva, A. M. Paganoni, D. Pigoli and V. Vitelli, ``Multivariate functional clustering for the morphological analysis of electrocardiograph curves''. \textit{Journal of the Royal Statistical Society: Series C (Applied Statistics)}, 62, 401-418, 2013. DOI: 10.1111/j.1467-9876.2012.01062.x

\bibitem{JP14A} J.~Jacques and C.~Preda, ``Model-based clustering of functional data.'', \textit{Computational Statistics and Data Analysis}, 71, 92--106, 2014. DOI:10.1016/j.csda.2012.12.004

\bibitem{JP14B} J.~Jacques and C.~Preda, ``Functional Data Clustering: A Survey.'', \textit{Advances in Data Analysis and Classification}, 8 (3), 231--255, 2014. DOI : 10.1007/s11634-013-0158-y.

\bibitem{JS03} G.~James and C.~Sugar, ``Clustering for sparsely sampled functional data''. \textit{Journal of the American Statistical Association}, 98, 397-408, 2003. DOI: 10.1198/016214503000189.

\bibitem{LXY00} B.~Liu, Y.~Xia  and P.S.~Yu, ``Clustering through decision tree construction''. \textit {CIKM 00. In: Proceedings of the ninth international conference on information and knowledge management}. ACM, New York, NY, USA, pp 20--29, 2000. DOI:10.1145/354756.354775

\bibitem{LP09} S.~L\'opez-Pintado and J.~Romo, ``On the concept of depth for functional data.'' \textit{Journal of the American Statistical Association}, 104, 718--734, 2009. DOI:10.1198/jasa.2009.0108


\bibitem{LSLG14} S.~L\'opez-Pintado, Y.~Sun, J.K.~Lin and M.G.~Genton,  ``Simplicial band depth for multivariate functional data''. \textit{Advances in Data Analysis and Classification}, 8, 321--338, 2014. DOI:10.1007/s11634-014-0166-6 


\bibitem{MLR14} B.~Martin-Barragan, R.~Lillo and J.~Romo, ``Interpretable support vector machines for functional data''. \textit{European Journal of Operational Research}, 232, 146.--155, 2014. DOI:10.1016/j.ejor.2012.08.017

\bibitem{M13} K.~Mosler, ``Depth Statistics''. \textit{Robustness and Complex Data Structures. Editors: Becker, C., Fried, R. and Kuhnt, S.} Springer Berlin Heidelberg. 17--34, 2013.

\bibitem{RHG09} J.~Ramsay, G.~Hooker and S.~Graves, \textit{Functional Data Analysis with R and Matlab}. Springer, New York, 2009.


\bibitem{RS05} J.~Ramsay and B.W.~Silverman, \textit{Functional Data Analysis} (2nd ed). Springer, New York, 2005.

\bibitem{SSVV10a} L.M.~Sangalli, P.~Secchi, S.~Vantini and V.~Vitelli, ``$k$-means alignment for curve clustering.'' \textit{Computational Statistics and Data Analysis}, 54, 1219--1233, 2010.  DOI:10.1016/j.csda.2009.12.008

\bibitem{SSVV10b} L.M.~Sangalli, P.~Secchi, S.~Vantini and V.~Vitelli,  ``Functional clustering and alignment methods with
applications''. \textit{Communications in Applied and Industrial Mathematics}, 1, 205--224, 2010. DOI: 10.1685/2010CAIM486.

\bibitem{SW05} N.~Serban and L.~Wasserman, ``CATS: Cluster Analysis by Transformation and Smoothing''. \textit{Journal of the American Statistical Association}, 100, 990--999, 2005.  DOI 10.1198/016214504000001574

\bibitem{SG11} Y.~Sun and M.G.~Genton, ``Functional boxplots''. \textit{Journal of Computational and Graphical Statistics}, 20, 316--334, 2011.  DOI: 10.1198/jcgs.2011.09224

\bibitem{TK03} T.~Tarpey and K.K.J.~Kinateder, ``Clustering functional data''. \textit{Journal of classification}, 20, 93--114, 2003. DOI : 10.1007/s11634-013-0158-y

\bibitem{TJ13} T.S.~Tian and G.M.~James, ``Interpretable dimension reduction for classifying functional data''. \textit{Computational Statistics and Data Analysis}, 57, 282--296, 2013. DOI:10.1016/j.csda.2012.06.017

\bibitem{TWH01} R.~Tibshirani, G.~Walther and T.~Hastie, ``Estimating the number of data clusters via the gap statistic''. \textit{Journal of the Royal Statistical Society, B}, 63, 411--423, 2001. DOI: 10.1111/1467-9868.00293.

\bibitem{TYI08} S.~Tokushige, H.~Yadohisa and K.~Inada, ``Crisp and Fuzzy k-means clustering algorithms for multivariate functional data''. \textit{Computational Statistics}, 21, 1--16, 2008. DOI:10.1007/s00180-006-0013-0

\bibitem{TS54} R.D.~Tuddenham and M.M.~Snyder, M.M., Physical growth of California boys and girls from birth to eighteen years, \emph{Tech. Rep. 1, University of CaliforniaPublications in Child Development.}, 1954.

\end{thebibliography}
\end{document}